\documentclass[a4paper,11pt]{article}

\usepackage{cite}
\usepackage{epsfig}

\newcommand{\fig}[4]{\begin{figure}[ht]\epsfxsize=#2\bigskip\centerline{\epsfbox{#1}}\caption{\small\it #3 \label{#4}}\bigskip\end{figure}}

\begin{document}

\title{A heuristic solution for the Pioneer anomaly employing an ELA metric with dark matter in the outskirts of the Solar system\\[2mm]}

\author{P. Castelo Ferreira\\[3mm] pedro.castelo.ferreira@ist.utl.pt\\[2mm] \small CENTRA--Instituto Superior T\'ecnico, Lisbon--Portugal\\ \small and\\ \small Eng. Electrot\'ecnica--U. Lus\'ofona Humanidades e Tecnologia, Lisbon--Portugal}

\date{}

\maketitle

\begin{abstract}
\noindent
It is discussed the possibility of describing the Doppler acceleration towards the Sun measured by the Pioneer space-craft above the heliocentric distance of $20\,AU$, known as the Pioneer anomaly, by considering an expanding background described by an expanding locally anisotropic (ELA) metric. This metric encodes both the standard local Schwarzschild gravitational effects and global universe expansion effects allowing simultaneously to fine-tune other gravitational effects
at intermediate scales, such as local dark matter and/or gravitational interactions corrections. Taking the Sun as central mass and the
heuristic ansatz for the dimensionless functional exponent parameter
$\alpha=3+\alpha_2(r_1)(1-U_{\odot})^2$ with a linear ansatz
$\alpha_2(r_1)=\alpha_2^{(1.0)}+\alpha_2^{(1.1)}\,r_1$ above $20\,AU$,
being $U_{\odot}$ the Schwarzschild gravitational potential of the Sun, the Pioneer effect is fully accounted for as due to the blue-shift of the
background space-time described by this metric. For compatibility with
orbital motion within the Solar system it is further considered a branch
ansatz for which $\alpha_2(r_1)=0$ below $20\,AU$. This construction has
the advantage of reducing the physical acceleration correction above $20\,AU$
down to $\sim 10^{-14}\,ms^{-2}$ outwards from the Sun, hence 4 orders of
magnitude below the measured Pioneer acceleration such that orbital corrections
to Neptune and Pluto are less notorious than when the full Pioneer acceleration is considered to be a gravitational acceleration. It is also discussed
the effective mass-energy density distribution above $20\,AU$, the
equation of state ($-1<\omega_r,\omega_\theta<-0.726$) due to the background corrections and the respective finite non-null total mass is computed by considering an upper cut-off for the
functional parameter $\alpha$. Such effective extended
configuration is temptatively  interpreted as a cold dark matter extended distribution.
\end{abstract}

%\begin{keyword}
%\pacs{95.35.+d}{Dark matter}
%\pacs{98.80.Jk}{Mathematical and relativistic aspects of cosmology}
%\pacs{04.20.-q}{Classical General Relativity}
%\end{keyword}

\section{Introduction}

The Pioneer space-crafts have been launch from Earth on 1972 and 1973 travelling outwards of the Solar system~[\citenum{Pioneer_1},\citenum{Pioneer_2}].
In between the heliocentric distances of $20\,AU$ and $70\,AU$, travelling at an approximately constant velocity of $v_p=12.2\times 10^3\,m\,s^{-1}$,
it was detected an unmodeled Doppler shift known as the Pioneer effect. If totally attributed to a physical acceleration, this effect corresponds to a radial acceleration (towards the Sun) of~[\citenum{Pioneer_2}]
\begin{equation}
a_p=-(8.74\pm 1.33)\times 10^{-10}\,m\,s^{-2}\ .
\label{a_p}
\end{equation}   

Although the origin for this anomaly is so far not established, many possible explanations for this effect have been considered in the literature, including modified gravity theories and models~[\citenum{mod_grav_p}], non-gravitational interactions~[\citenum{extra_matter}], extra-dimensional theories and models~[\citenum{extra_dimensions}] and dark matter~[\citenum{dm_p}].

The main purpose of the present work is to investigate whether the Pioneer effect can be consistently obtained due to a background
described by the ELA metric~[\citenum{PLB,e-print}], an ansatz that interpolates between the Schwarzschild (SC) metric~[\citenum{Schwarzschild}]
near massive bodies and the Robertson-Walker (RW) metric~[\citenum{RW}] at spatial infinity, hence describing local 
matter in an expanding Universe~[\citenum{Hubble}]. The shift function of this metric can be fine-tuned through an exponential parameter allowing to describe intermediate scale gravitational effects~[\citenum{e-print,curves}] which may be interpreted
either as due to local cold dark matter (CDM) effects~[\citenum{DM}] or modified theories of gravity~[\citenum{mod_grav}]. Although this fine-tuning is technically more challenging than semi-classical heuristic approaches, has the following advantages:
\begin{itemize}
\item it is a covariant formulation allowing to deal with CDM within the framework of General Relativity;
\item in particular allows for direct covariant computations of gravitational accelerations and gravitational red-shifts (blue-shifts);
\item allows as well for QFT cross-section computations in curved backgrounds to include CDM effects (a metric is required to define the Dirac equation);
\item it allows for the description of intermediate scale CDM effects maintaining the local and global known physical properties of space-time.
\end{itemize}

In the following we will employ two distinct coordinate systems, expanding coordinates, $r_1$, for which the area of the sphere is $A=4\pi\,r_1^2$
such that the coordinate quantities correspond directly to
measurable lengths and the standard Robertson-Walker coordinates, $r$, for which the area of the sphere is $A=4\pi\,a^3r^2$ being suitable for direct computations of measurable red-shifts~[\citenum{e-print}]. The map between both these coordinate systems is $r_1=a\,r$ and the ansatz for the ELA metric is~[\citenum{PLB,e-print}] 
\begin{equation}
\begin{array}{rcl}
ds^2&=&\displaystyle(1-U)(cdt)^2-\left(\frac{dr_1}{\sqrt{1-U}}-\frac{H\,r_1}{c}(1-U)^{\frac{\alpha}{2}}(cdt)\right)^2-d\Omega_1^2\\[3mm]
&=&\displaystyle(1-U)(cdt)^2-a^2\left(\frac{dr}{\sqrt{1-U}}+\frac{H\,r}{c}\left(1-(1-U)^{\frac{\alpha}{2}}\right)(cdt)\right)^2-a^2d\Omega^2\ ,
\end{array}
\label{ELA_metric}
\end{equation}
where $a$ is the universe scale factor, $H=\dot{a}/a$ is the time dependent Hubble rate, $U=2GM/(c^2\,r_1)=2GM/(c^2\,ar)$ is the standard SC gravitational potential, $M$ is the standard SC mass, 
$c$ is the speed of light, $G$ is the Newton gravitational constant and the solid angle line-elements are $d\Omega_1^2=r_1^2(d\theta^2+\sin^2\theta d\varphi^2)$ and $d\Omega^2=r^2(d\theta^2+\sin^2\theta d\varphi^2)$.

This ansatz generalizes previous ansatz and solutions in the literature~[\citenum{McV}] allowing for local anisotropy~[\citenum{anisotropy}] and having the novelty of describing a
space-time without singularities at the SC radius simultaneously maintaining
as asymptotic limits the SC metric near the SC event horizon and
the RW metric at spatial infinity. To ensure these properties of space-time the following limits must be obeyed~[\citenum{PLB,e-print}]
\begin{equation}
\begin{array}{ll}
\displaystyle\lim_{r_1\to 0}(1-U)^\alpha=0&\mathrm{:\ SC\ mass\ pole\ \mathit{M}\ is\ maintained}\ ,\\[5mm]
\displaystyle\lim_{r_1\to r_{1.SC}}\alpha\ge 3&\mathrm{:\ space-time\ is\ singularity\ free\ at\ SC\ horizon}\ ,\\[5mm]
\displaystyle\lim_{r_1\to \infty}(1-U)^\alpha= 1&\mathrm{:\ converges\ to\ RW\ metric}\ .
\end{array}
\label{alpha_lim}
\end{equation}
In the following we consider as ansatz for the exponential parameter
$\alpha$ a second order expansion in the gravitational field given by
\begin{equation}
\alpha(r_1)=\alpha_0(r_1)+\alpha_2(r_1)(1-U)^2\ \ ,\ \alpha_0(r_1)=3\ ,\ \alpha_2(r_1\to 0)<0.
\label{alpha}
\end{equation}
This ansatz allows for higher values of the corrections over intermediate scales than a first order expansion ansatz in the gravitational field maintaining the asymptotic limits expressed in~(\ref{alpha_lim}). In particular, in the remaining of this work, we will heuristically fit the functional parameter $\alpha_2$ to the experimentally measured Pioneer effect which is fully accounted as a background blue-shift and compute a finite non-null mass for the extended correction to the background which may be interpreted as a CDM extended distribution.

\section{Describing the Pioneer Effect\label{sec.Pioneer}}

The Pioneer effect was measured using a two-way Doppler shift~[\citenum{Pioneer_1}]. The standard gravitational red-shift
as described by the SC metric is null, as for the standard cosmological red-shift is negligible as noted in~[\citenum{Pioneer_2,e-print}] being below the accuracy of the Pioneer measurements.
For this reason the Pioneer effect is often attributed to a correction to the gravitational acceleration~[\citenum{mod_grav_p,extra_dimensions,dm_p}].

Taking as central mass the Sun and neglecting the contribution of other bodies in the solar system, the corrections to the gravitational acceleration felt by a test mass travelling in the background given by the ELA metric~(\ref{ELA_metric}) with respect to the General Relativity gravitational
acceleration corresponding to a Ricci-flat background described by the SC metric 
is~[\citenum{e-print,curves}]
\begin{equation}
\begin{array}{rcl}
\delta\ddot{r}_p&\approx&\displaystyle\frac{H^2r_p}{2}\,\left(1-U_{\odot}\right)^{\alpha}\left(2(1-U_{\odot})-(1+\alpha)r_p\,U_{\odot}'\right.\\[5mm]
&&\displaystyle\left.-2(1+q)(1-U_{\odot})^{\frac{1}{2}-\frac{\alpha}{2}}+r_p(1-U_{\odot})\ln(1-U_{\odot})\alpha'\right)+O(H^4)\ ,
\end{array}
\label{F}
\end{equation}    
where the primed quantities stand for differentiation with respect to the radial coordinate $r_1$, $r_p$ stands for the geometrical (measurable) distance from the Sun to the Pioneer space-craft and $U_{\odot}=2GM_{\odot}/(c^2r_p)$ stands for the standard SC gravitational potential of the Sun with
$M_{\odot}=1.99\times 10^{30}\,Kg$.

In addition there is a correction to the gravitational red-shift. The measurable
red-shift due to the background is obtained from the line element~(\ref{ELA_metric}). For a light ray travelling approximately at a radial trajectory with respect to the Sun described by the four-vector $k^\mu\approx(\omega/c,k^r,0,0)$ , from earth at $r^{(0)}_1=a_0r^{(0)}\approx 1\,AU$ to the Pioneer space-craft at $r^{(1)}_1=a_1\,r^{(1)}=r_p$ and back to earth at $r^{(2)}_1=a_2\,r^{(2)}\approx 1\,AU$ we obtain that the ratios
between the radial wave numbers of the emitted radiation from earth
$k^r_{(0)}$, the radiation received at the Pioneer space-craft $k^r_{(1)}$
and the radiation received back at Earth $k^r_{(2)}$ are
\begin{equation}
\frac{k^r_{(1)}}{k^r_{(0)}}=\frac{a_0}{a_1}\,\frac{1-U_p-\delta}{1-U_e}\ ,\ \frac{k^r_{(2)}}{k^r_{(1)}}=\frac{a_1}{a_2}\,\frac{1-U_e}{1-U_p+\delta}\ ,\ \delta=\frac{H\,r_p}{c}\left(1-(1-U)^{\frac{\alpha}{2}+\frac{1}{2}}\right)\ ,
\end{equation}
where $U_p$ and $U_e$ are the gravitational potential of the Sun evaluated
at the Pioneer space-craft and at Earth, and $a_0$, $a_1$ and $a_2$ stand for
the Universe scale factor $a=a(t)$ evaluated at the time of emission of the radiation from Earth, the time of receiving and re-emission at the Pioneer space-craft and the time of receiving it back at Earth, respectively.
Further noting that expanding the
Universe scale factor to first order in the Hubble rate we obtain that
$a_1=a_0(1+(H/c)r_p)+O(H^2)$ and $a_2=a_0(1+2(H/c)r_p)+O(H^2)$, the measurable
red-shift (or blue-shift, if negative) due to the background for a two
way Doppler measurement from Earth to Pioneer and back is
\begin{equation}
\frac{\Delta\nu}{\nu_0}=\frac{k^r_{(1)}}{k^r_{(0)}}\,\frac{k^r_{(2)}}{k^r_{(1)}}-1=\frac{1}{1+2\frac{H\,r_p}{c}}\,\frac{1-U_p-\delta}{1-U_p+\delta}-1+O(H^4)
\end{equation}
Hence the contribution to the Pioneer non-physical acceleration due
to the background red-shift is obtained by differentiating the standard
two way Doppler shift $\Delta\nu/\nu_0=-2\dot{r}_p/c$
\begin{equation}
\delta\ddot{r}_p^{Doppler}=-c\,\frac{v_p}{2}\,\left(\frac{\Delta\nu}{\nu_0}\right)'\ ,
\label{F_Doppler}
\end{equation}
where the prime denotes differentiation with respect to the radial geometric distance (the coordinate $r_1$), $v_p=\dot{r}_p$ for the radial geometric
velocity of the space-craft (obtained from the differentiation $\partial_{t}=\partial_tr_1\,\partial_{r_1}$) and we have neglected relativistic
corrections ($\gamma\approx 1$ for $v_p\ll c$).

Hence, if the Pioneer acceleration is totally due to the background encoded
in the metric~(\ref{ELA_metric}), it is the sum of the contribution~(\ref{F}) and (\ref{F_Doppler})
\begin{equation}
a_p=\delta\ddot{r}_p+\delta\ddot{r}_p^{Doppler}\ .
\end{equation}
Assuming $\alpha_0(r_1)=3$ and an approximately linear functional exponent $\alpha_2(r_1)=\alpha^{(0)}_2+\alpha^{(1)}_2\,r_1$ for radial distances above $r_1=20\,AU$ the Pioneer data is fitted, within the error bars, for the
values of the constant coefficients $\alpha^{(0)}_2=-1.23\times 10^9$
and $\alpha^{(1)}_2=-7.12\times 10^{-3}\,m^{-1}$. Fits within half of the error bars
are also straight forwardly obtained as pictured in figure~\ref{fig.1}. Within the range $r_p\in[20,70]\,AU$ the dominant
contribution to $a_p$ is the background blue-shift, $\delta\ddot{r}_p^{Doppler}\sim -10^{-10}\,ms^{-2}$. As for
the value of the physical acceleration, it is 4 orders of magnitude below this value, $\delta\ddot{r}_p\sim +10^{-14}\,ms^{-2}$ (outwards from the Sun).
\fig{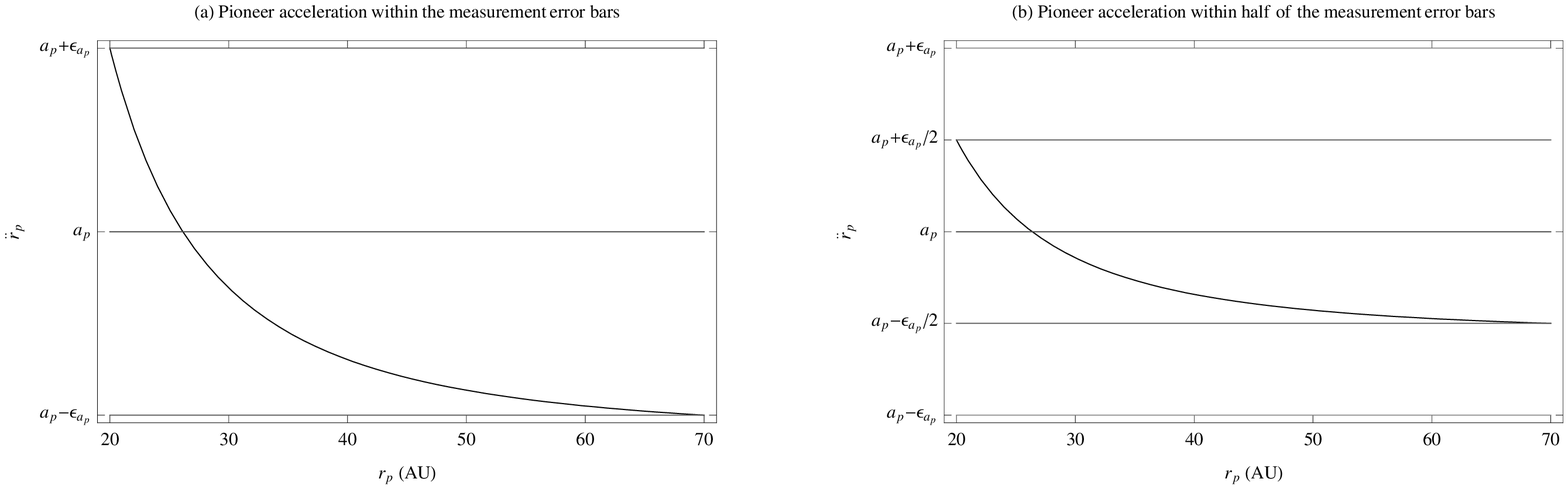}{130mm}{Fit of the linear coefficient $\alpha_2(r_p)=\alpha^{(0)}_2+\alpha^{(1)}_2\,r_p$ to the Pioneer acceleration $a_p$~(\ref{a_p}): (a) within the measurement error bars, $\alpha^{(0)}_2=-1.225\times 10^9$  and $\alpha^{(1)}_2=-7.118\times 10^{-3}\,m^{-1}$; (b) within half of the measurement error bars, $\alpha^{(0)}_2=-9.459\times 10^8$  and $\alpha^{(1)}_2=-7.067\times 10^{-3}\,m^{-1}$.}{fig.1}

\section{Compatibility with planetary motion}
Assuming the linear fit just discussed, within the Solar system, we obtain that the physical acceleration is positive above $r_1=12.097\,AU$ increasing monotonically (at
$r_1=70\,AU$ its value is $\delta\ddot{r}_p=8.626\times 10^{-14}\,ms^{-2}$) and it is negative below $r_1=12.097\,AU$ overcoming the Newton acceleration below $r_1=0.890\,AU$. These results are clearly not verified experimentally, the orbital motion within the solar system is driven and approximately described to a very good accuracy by the Newton acceleration. This problem is straight forwardly solved by considering a cut-off
at $r_1=20\,AU$ such that, below $r_1=20\,AU$, this coefficient is approximately null. Hence we obtain the following heuristic ansatz
\begin{equation}
\alpha_2(r_1)=
\left\{\begin{array}{lcl}
-\alpha_2^{(0.0)}&,&r_1\in[0,20[\,AU\\[5mm]
\alpha_2^{(1.0)}+\alpha_2^{(1.1)}\,r_1&,& r_1\in[20,R_{\mathrm{Max}}]\,AU
\end{array}\right.
\label{Spline_1}
\end{equation}
with $0<\alpha_2^{(0.0)}\ll 1$ being an arbitrarily small non-null positive constant that ensures that the limits~(\ref{alpha_lim}) are obeyed~[\citenum{PLB}], $\alpha_2^{(1.0)}=-1.225\times 10^9$ and $\alpha_2^{(1.1)}=-7.118\times 10^{-3}\,m^{-1}$. We have also considered an upper cut-off $R_{\mathrm{Max}}$ that will be discussed in the next section. Although not a very elegant solution this heuristic ansatz allows to describe the Pioneer effect maintaining compatibility with the planetary
motion within the solar system.

Nevertheless the effects of this correction to the gravitational
acceleration should be measurable at least for Pluto and Neptune: as an example, for the ansatz~(\ref{Spline_1}), if this
correction is fully accounted by an orbital radius variation~[\citenum{e-print}]
we would obtain, respectively,
$\dot{r}_{orb}/r_{orb}\approx 2.62\times 10^{-26}\,s^{-1}$ and
$\dot{r}_{orb}/r_{orb}\approx 1.21\times 10^{-26}\,s^{-1}$ which is
well above the experimentally measured orbital radius variation for
inner planets in the Solar system such as Venus and Mars, $|\dot{r}_{orb}/r_{orb}|< 10^{-40}\,s^{-1}$~[\citenum{Uzan}].
However further analytical or numerical results are required
to exactly predict the effects on planetary motion due to this correction~[\citenum{e-print}].

\section{Mass-Energy density and Possible Dark Matter Interpretation}

So far we have not analysed neither the mass-energy density neither the equation of state for the ELA metric~(\ref{ELA_metric}). The mass-energy density ($\rho$) and anisotropic
pressures ($p_r$ and $p_\theta=p_\varphi$) are
\begin{equation}
\begin{array}{rcl}
\rho&=&\displaystyle\frac{H^2}{8\pi\,G}\left(1-U\right)^{\alpha-1}\left(\alpha\,U+(1-U)\left(3+r_1\log\left(1-U\right)\partial_{r_1}\alpha\right)\right)\ ,\\[5mm]
p_r&=&\displaystyle\frac{H^2}{8\pi\,G}\left(1-U\right)^{\frac{\alpha-1}{2}}\left(2(1+q)-\left(1-U\right)^{\frac{\alpha-1}{2}}\left(\alpha\,U\right.\right.\\[3mm]
&&\displaystyle\hfill\left.+(1-U)(3+r_1\log\left(1-U\right)\partial_{r_1}\alpha\right)\bigg)\ ,\\[5mm]
p_\theta&=&\displaystyle\frac{H^2}{16\pi\,G}\left(1-U\right)^{\frac{\alpha-3}{2}}\Bigg[4(1+q)\left(1-U\left(1-\frac{\alpha}{4}\right)\right)\\[3mm]
&&\displaystyle-2\left(1-U\right)^{\frac{\alpha-1}{2}}\left(3-3U\left(2-U\left(1-\frac{\alpha}{2}\right)\right)\left(1-\frac{\alpha}{3}\right)+r_1U(1-U)\partial_{r_1}\alpha\right)\\[3mm]
&&\displaystyle+r_1(1-U)\log(1-U)\left((1+q)\partial_{r_1}\alpha-2U\left(1-U\right)^{\frac{\alpha-1}{2}}\alpha\partial_{r_1}\alpha\right.\\[3mm]
&&\displaystyle\left.\left(1-U\right)^{\frac{\alpha+1}{2}}\left(6\partial_{r_1}\alpha+r_1\log(1-U)(\partial_{r_1}\alpha)^2+r_1\partial^2_{r_1}\alpha\right)\right)\Bigg].
\end{array}
\label{rho}
\end{equation}
We note that, for the linear ansatz with one single branch for all values of
the radial coordinate discussed in section~\ref{sec.Pioneer},
$\alpha_2=\alpha_{2}^{(1.0)}+\alpha_{2}^{(1.1)}\,r_1$,
the mass-energy is negative below $r_1=8.06\,AU$
hence violating causality. The branch ansatz~(\ref{Spline_1}) also solves
this problem, for this ansatz $\rho>0$ for all values of the radial coordinate above the SC radius $r_1>r_{1.\mathrm{SC}}=2GM/c^2$. 

When the limits~(\ref{alpha_lim}) are obeyed~[\citenum{PLB}], at $r_1=r_{1.\mathrm{SC}}$, $\rho$ coincides with
the mass-energy density for Ricci flat space-times, being null $\rho(r_1\to r_{1.\mathrm{SC}})=0$, and at spatial infinity it asymptotically converges to
the RW mass-energy density $\rho(r_1\to +\infty)=\rho_{RW}=3\,h^2/(8\pi\,G)$, as well as $p_{r}(r_1\to +\infty)=p_{\theta}(r_1\to +\infty)=p_{\varphi}(r_1\to +\infty)=p_{RW}=(2q-1)\,h^2/(8\pi\,G)$ such that at spatial infinity the RW equation of state is obtained $\omega_{RW}=(2q-1)/3\approx -0.726$ (this value is computed within the $\Lambda$CDM model assuming that the pressure of the background is given only by the cosmological constant~[\citenum{e-print}]). Hence this mass-energy distribution has been interpreted in~[\citenum{PLB}], for a constant functional
exponent $\alpha$ as a local anisotropic extended correction to the global isotropic cosmological background due to a local point mass vanishing at spatial infinity. In addition, when considering a radial coordinate dependence on the functional parameter $\alpha_2$, it has been put forward in~[\citenum{e-print,curves}] that the ELA metric~(\ref{ELA_metric}) may also
encode the corrections to the expanding background due to the existence of a local CDM distribution. If this is the case the total CDM mass can be computed corresponding to the mass of such extended corrections to the cosmological background. Specifically the CDM mass within a shell of radius $R_1$ for a continuous ansatz is
\begin{equation}
M_{CDM}(R_1)=4\pi\int_{0}^{R_1}r_1^2(\rho-\rho_{RW})dr_1=\frac{H^2}{2\,G}\left(\left(1-U(R_1)\right)^{\alpha(R_1)}-1\right)R_1^3\ .
\end{equation}
As long as the limits~(\ref{alpha_lim}) are obeyed $M_{CDM}(R_1\to 0)=0$. For convergence of this quantity at spatial infinity more severe conditions than the limits~(\ref{alpha_lim}) are required, in particular it can be fine-tuned either to be null (e.g. for $\alpha(r_1\to +\infty)={\mathrm{const}}/r_1^n,\,n>2$) or to a multiple of the central mass (e.g. for $\alpha(r_1\to +\infty)={\mathrm{const}}/r_1^2$ we obtain $M_{CDM}(R_1\to +\infty)=H^2{\mathrm{const}}\,M_\odot/(2G)$). However for the ansatz~(\ref{Spline_1}) for $\alpha_2$,
$M_{CDM}(R_1\to +\infty)=+\infty$ and there is a discontinuity at $r=20\,AU$ contributing a finite non-null contribution to the total mass.

Further recalling that in average the ratio of CDM to baryonic matter is $\Omega_{cdm}/\Omega_{b}=5$~[\citenum{wmap}],
if the ELA metric corrections are to be interpreted as due to a CDM distribution, each individual body should have a non-null finite value for the quantity $M_{CDM}$ contributing to
$\Omega_{cdm}$. This is straight forwardly achieved by considering a non-continuous upper cut-off for $\alpha$ at some value of the radial coordinate $r_1=R_{\mathrm{Max}}$ such that above this value $\alpha=\alpha_0=\alpha_2=0$ and the metric for large radii coincides with the original isotropic McVittie metric~[\citenum{McV}] for which $\rho=\rho_{RW}$. The CDM mass of this extended configuration is given by $M_{CDM}(R_{\mathrm{Max}})+M_{CDM}(20^-\,AU)-M_{CDM}(20^+\,AU)\approx M_{CDM}(R_{\mathrm{Max}})$ and the cut-off $R_{\mathrm{Max}}$ corresponds to the edge of the CDM extended distribution. For the Sun, assuming the ratio $M_{CDM}/M_{\odot}=5$, we obtain $R_{\mathrm{Max}}= 38317\,AU$. We remark that this upper cut-off
also ensures that isotropy is exactly retrieved above $R_{\mathrm{Max}}$.

As for the equation of state for the ELA metric, we have already notice that at spatial infinity it coincides with the standard
RW equation, hence neither the local baryonic matter correction to the background, neither the local CDM matter correction to the background modify the global equation of state: statistically
the respective pressures are null $p_b=p_{cdm}=0$, only the
masses contribute to the global background densities $\Omega_b$
and $\Omega_{cdm}$. However
we do note that locally there are deviation from the global
densities, pressures and equation of state. For baryonic matter
with a constant exponent $\alpha$, the equation of state was analysed in detail in~[\citenum{PLB}]. As for the
extended distribution in the range $r\in[20\,AU,R_{\mathrm{max}}]$ corresponding to the ansatz~(\ref{Spline_1}),
we obtain at $r=20AU$, $\rho=2.71\times 10^9\,\rho_{RW}$, $p_r=3.74\times 10^9\,p_{RW}$ and $p_\theta=p_\varphi=2.74\times 10^9\,p_{RW}$ such that $\omega_r=p_r/\rho=-1+6.81\times 10^{-6}$ and $\omega_\theta=\omega_\varphi=p_\theta/\rho=-7.33$. These
values change monotonically up to $r=R_{\mathrm{max}}$ where we obtain $\rho=1.36\times 10^9\,\rho_{RW}$, $p_r=1.87\times 10^9\,p_{RW}$ and $p_\theta=p_\varphi=1.87\times 10^9\,p_{RW}$ such that $\omega_r=-1+7.44\times 10^{-6}$ and $\omega_\theta=\omega_\varphi=-1+2.18\times 10^{-4}$.
We further remark that the values of the equation of state, $-1<w_r,\,w_\theta<-0.726$, indicate that this construction may be compatible with a locally anisotropic configuration of scalar fields~[\citenum{scalar}].

Again we note that, although a clearly unelegant heuristic solution not derived from first principles, this construction allows for the interpretation of
the Pioneer effect as due to dark matter, as well as, generally,
allowing for a covariant description of local dark matter maintaining compatibility with today's global parameters of the universe.

\section{Conclusions}

In this work we have considered a linear ansatz for the functional
parameter $\alpha_2$ of the ELA metric~(\ref{ELA_metric}) which allows
to describe the Pioneer effect. The main contribution is due to the
background gravitational blue-shift such that the physical acceleration
contribution to the Pioneer acceleration is only of order $\sim 10^{-14}\,ms^{-2}$,
hence 4 orders of magnitude below the measured Pioneer acceleration~[\citenum{Pioneer_1,Pioneer_2}]. 

Within this framework are also evaluated the orbital radius variation $\dot{r}/r$ for
Pluto and Neptune. Orbital motion corrections to these planets are a common
consequence of any gravity theory or model that allows to describe the Pioneer effect~[\citenum{planets}].
However we remark that the construction suggested in this work
allows to significantly decrease the magnitude of such corrections due to the Pioneer effect
being mostly attributed to a background blue-shift instead of a correction to the gravitational acceleration.

In addition we also discuss the possibility of interpreting this effect as due
to a CDM extended distribution from $r_1=20\,AU$ up to a cut-off $r_1=R_{\mathrm{Max}}$ with a computable fine-tuneable total mass $M_{CDM}$.\\

\noindent{\sc\bf Acknowledgments}:
Work supported by grant SFRH/BPD/34566/2007 from FCT-MCTES.


\begin{thebibliography}{99}

\bibitem{Pioneer_1} J. D. Anderson et al., \textit{Indication, from Pioneer 10 / 11, Galileo, and Ulysses data, of an apparent anomalous, weak, long range acceleration}, Phys. Rev. Lett. {\bf 81} (1998) 2858-2861, \texttt{gr-qc/9808081};
\textit{Study of the Anomalous Acceleration of Pioneer 10 and 11}, Phys. Rev. {\bf D65} (2002) 082004, \texttt{gr-qc/0104064};
\textit{The search for a standard explanation of the Pioneer anomaly}, Mod. Phys. Lett. {\bf A17} (2002) 875-886, \texttt{gr-qc/0107022}.

\bibitem{Pioneer_2} S. G. Turyshev and V. T. Toth, \textit{The Pioneer Anomaly}, Living Rev. Rel. {\bf 13} (2010) 4, \texttt{arXiv:1001.3686}.

\bibitem{mod_grav_p} Dag {\O}stvang, \textit{An explanation of the 'Pioneer effect' based on quasi-metric relativity}, Class. Quant. Grav. {\bf 19} (2002) 4131-4140, \texttt{gr-qc/9910054};
M.-T. Jaekel and S. Reynaud, \textit{Gravity tests in the solar system and the Pioneer anomaly}, Mod. Phys. Lett. {\bf A20} (2005) 1047-1055, \texttt{gr-qc/0410148};
J. R. Brownstein and J. W. Moffat, \textit{Gravitational solution to the Pioneer 10/11 anomaly}, Class. Quant. Grav. {\bf 23} (2006) 3427-3436, \texttt{gr-qc/0511026};
M. E. McCulloch, \textit{Modelling the Pioneer anomaly as modified inertia}, Mon. Not. Roy. Astron. Soc. {\bf 376} (2007) 338-342, \texttt{astro-ph/0612599};
R. Saffari, S. Rahvar, \textit{f(R) Gravity: From the Pioneer Anomaly to the Cosmic Acceleration}, Phys. Rev. {\bf D77} (2008) 104028;
M. N. Smolyakov, \textit{Gravity in Brans-Dicke theory with Born-Infeld scalar field and the Pioneer anomaly}, Int. J. Mod. Phys. {\bf A25} (2010) 1135-1145,\texttt{arXiv:0907.3744}.

\bibitem{extra_matter} J. Wood, W. Moreau, \textit{Solutions of Conformal Gravity with Dynamical Mass Generation in the Solar System}, \texttt{gr-qc/0102056};
S. Capozziello, S. De Martino, S. De Siena and F. Illuminati, \textit{Non-Newtonian Gravity, Fluctuative Hypothesis and the Sizes of Astrophysical Structures}, Mod. Phys. Lett. {\bf A16} (2001) 693-706, \texttt{gr-qc/0104052};
R. Foot and R. R. Volkas, \textit{A Mirror world explanation for the Pioneer spacecraft anomalies?}, Phys. Lett. {\bf B517} (2001) 13-17, \texttt{hep-ph/0108051};
M. M. Nieto, S. G. Turyshev and J. D. Anderson, \textit{Directly measured limit on the interplanetary matter density from Pioneer 10 and 11}, Phys. Lett. {\bf B613} (2005) 11-19, \texttt{astro-ph/0501626};
W. F. Hall, \textit{Can charge drag explain the Pioneer anomaly?}, Phys. Lett. {\bf B646} (2007) 1-5.

\bibitem{extra_dimensions} W. B. Belayev, \textit{Five-dimensional Gravity and the Pioneer Effect}, \texttt{gr-qc/0209095};
O. Bertolami and J. Par\'amos, \textit{The Pioneer anomaly in a bimetric theory of gravity on the brane}, Class. Quant. Grav. {\bf 21} (2004) 3309-3321, \texttt{gr-qc/0310101}.

\bibitem{dm_p} F. Munyaneza and R. D. Viollier, \textit{Heavy neutrino dark matter in the solar system}, \texttt{astro-ph/9910566};
E. Schmutzer, \textit{Distribution of dark matter around a central body, pioneer effect and fifth force}, \texttt{gr-qc/0106049}.

\bibitem{PLB} P. Castelo Ferreira, Phys. Lett. {\bf B 684} (2010) 73-76.

\bibitem{e-print} P. Castelo Ferreira, unpublished, \texttt{arXiv:0907.0847}.

\bibitem{Schwarzschild} K. Schwarzschild, Sitzungsber. Preuss. Akad. Wiss. Berlin - Math. Phys. (1916) 189-196, \texttt{physics/9905030};Sitzungsber. Preuss. Akad. Wiss. Berlin - Math. Phys. (1916) 424-434, \texttt{physics/9912033}.

\bibitem{RW} G. Lema\^itre, M. N. {\bf 91} (1931) 490-501;
H. P. Robertson, Astr. J. {\bf 82} (1935) 248-301; {\bf 83} (1936) 187-201; 257-271;\\ A. G. Walker, Proc. London Math. Soc. {\bf 42} (1936) 90-127.

\bibitem{Hubble} E. P. Hubble, Proc. Nat. Acad. Sci. U.S. {\bf 15} (1929) 169-173; A. Sandage, Astrophys. J. {\bf 136} (1962) 319.

\bibitem{curves} P. Castelo Ferreira, \texttt{arXiv:1006.1619}, the metric in this e-print is wrong and for the correct metric there are regions within the galaxy plane with strictly negative mass-energy. Currently being rewritten.

\bibitem{DM} F. Zwicky, Hlv. Phys. Acta {\bf 6} (1933) 110-127; K. A. Olive, \texttt{astro-ph/0301505}; D. Hooper, \texttt{arXiv:0901.4090}.

\bibitem{mod_grav}  M. Pilgrim, \textit{A Modification of the Newtonian Dynamics as a Possible Alternative to the Hidden Mass Hypothesis}, Astrophys. J. {\bf 270} (1983) 365-370;
J. W. Moffat, \textit{Gravitational Theory, Galaxy Rotation Curves and Cosmology Without Dar Matter}, JCAP 0505 (2005) 003, \texttt{astro-ph/0412195};
\textit{Scalar-Tensor-Vector Gravity Theory}, JCAP 0603 (2006) 004, \texttt{gr-qc/0506021}.

\bibitem{McV} G. C. McVittie, Mon. N. Roy. A. Soc. {\bf 93} (1933) 325;
M. Ferraris, M. Francaviglia and A. Spallicci, Nuovo Cimento {\bf B 111} (1996) 1031-1036;
M. Mizony and M. Lachi\`eze-Rey, Astron. Astrophys. {\bf 434} (2005) 45-52, \texttt{gr-qc/0412084};
G. S. Atkins, J. McDonnell and R.N. Fell, \texttt{gr-qc/0612146}. 

\bibitem{anisotropy} M. Davis and P. J. E. Peebles, Ann. Rev. Astron. Astrophys. {\bf 21} (1983) 109-130.

\bibitem{Uzan} J.-P. Uzan, Rev. Mod. Phys. {\bf 75} (2003) 403, \texttt{hep-ph/0205340}.

\bibitem{wmap} E. Komatsu et al., Astr. J. Supp. {\bf 180} (2009) 330-376, \texttt{arXiv:0803.0547}.

\bibitem{scalar} I. Zlatev, L. Wang and P. J. Steinhardt, \textit{Quintessence, Cosmic Coincidence, and the Cosmological Constant}, Phys. Rev. Lett. {\bf 82} (1999) 896-899, \texttt{astro-ph/9807002}; V. Sahni and L. Wang, \textit{A New Cosmological Model of Quintessence and Dark Matter}, Phys. Rev. {\bf D62} (2000) 103517, \texttt{astro-ph/9910097}; W. Zimdahl and D. Pavon, \textit{Interacting quintessence}, Phys. Lett. {\bf B521} (2001) 133-138, \texttt{astro-ph/0105479}.

\bibitem{planets} L. Iorio and G. Giudice, \textit{What do the orbital motions of the outer planets of the Solar System tell us about the pioneer anomaly?}, New Astron. {\bf 11} (2006) 600-607, \texttt{gr-qc/0601055};
Kjell Tangen, \textit{Could the Pioneer anomaly have a gravitational origin?}, Phys. Rev. {\bf D76} (2007) 042005, \textit{gr-qc/0602089};
L. Iorio, \textit{Can the Pioneer anomaly be induced by velocity-dependent forces? Tests in the outer regions of solar system with planetary dynamics}, Int. J. Mod. Phys. {\bf D18} (2009) 947-958, \texttt{arXiv:0806.3011}.

\end{thebibliography}
\end{document}